\documentclass[referee]{raa}           % ADD COMMENTS ABOUT FLUX DENSITIES ETC.

\usepackage{graphicx,times}
\usepackage{natbib}
\usepackage{amssymb,amsmath}
\usepackage{longtable}
\bibpunct{(}{)}{;}{a}{}{,}

\begin{document}

   \title{Flux density measurements for 32 pulsars in the 20\,cm observing band}
 \volnopage{ {\bf 2012} Vol.\ {\bf X} No. {\bf XX}, 000--000}
   \setcounter{page}{1}

   \author{Yan-Wei Xie\inst{1,2}, Jingbo Wang\inst{3,4}, George Hobbs\inst{5,2}, Jane Kaczmarek\inst{5}, Di Li\inst{2,6}, Jie Zhang\inst{1,7}, Shi Dai\inst{5}, Andrew Cameron\inst{5}, Lei Zhang\inst{2,5}, Chenchen Miao\inst{2,6}, Mao Yuan\inst{2,6}, Shen Wang\inst{2,6,8}, Songbo Zhang\inst{9,5}, Heng Xu\inst{10,11,2}, Renxin Xu\inst{10,11}
   }
%% For single author or all the authors from an institute, use "\inst{}" only

   \institute{ School of Physics and Space Science, China West Normal University, Nanchong 637002, China;  \\
%% Please give the E-mail address of the author, to whom future correspondence and
        \and
             National Astronomical Observatories, Chinese Academy of Sciences, Datun Road, Chaoyang District, Beijing 100012, China
\\
\and
Xinjiang Astronomical Observatory, Chinese Academy of Sciences, 150 Science 1-Street, Urumqi,
Xinjiang 830011, China {\it wangjingbo@xao.ac.cn}
\\
\and
Key Laboratory of Radio Astronomy, Chinese Academy of Sciences, 150 Science 1-Street, Urumqi, Xinjiang, 830011, China
\\
	\and
CSIRO Astronomy and Space Science, PO Box 76, Epping, NSW 1710, Australia
\\
\and
CAS Key Laboratory of FAST, NAOC, Chinese Academy of Sciences, Beijing 100101, China
\and
College of Physics and Electronic Engineering, Qilu Normal University, Jinan, 250014, China
\and
Astronomy Department, Cornell University, Ithaca, NY 14853, USA
\and
Purple Mountain Observatory, Chinese Academy of Sciences, Nanjing 210008, China
\and
Kavli Institute for Astronomy and Astrophysics, Peking University, Beijing 100871, China
\and
Department of Astronomy, School of Physics, Peking University, Beijing 100871, China
\vs \no
%   {\small Received ; accepted }
}

\abstract{Flux densities are fundamental observational parameters that describe a pulsar.  In the current pulsar catalogue,  27\% of the listed radio pulsars have no flux density measurement in the 20\,cm observing band. Here, we present the first such measurements for 32 pulsars observed using the Parkes Radio Telescope. We have used both archival and new observations to make these measurements. Various schemes exist for measuring flux densities and we show how the measured flux densities vary between these methods and how the presence of radio-frequency interference will bias the flux density measurements.
}

   \authorrunning{Y.-W. Xie et al. }           
   \titlerunning{Flux density measurements for 32 pulsars in the 20\,cm band}  % title_head in odd pages
   \maketitle

\section{Introduction}          
\label{sect:intro}

Measurement of a pulsar's flux density in one or more observing bands provide information on the pulsar emission process (for instance, through measurements of the spectral index of the emission; e.g., \citealt{kbl+17}) as well as providing a means to probe the interstellar medium (e.g., \citealt{kcs+13,lmj+16, kcw+18}). As almost all the major radio observatories carry out observations in the 20\,cm observing band, knowledge of pulsar flux densities in this band is essential. Such information is used to predict how many pulsars a new survey may discover, to explain why some pulsars are not detected in a particular observation and to select key, bright pulsars to observe for commissioning or calibration observations.   

The Australia Telescope National Facility (ATNF) pulsar catalogue (\citealt{2005yCat.7245....0M}) provides a repository for pulsar observational parameters.   Out of the 2659 radio pulsars in the catalogue (we have used the most up-to-date version, 1.59), 704 do not have a flux density determination in the 20\,cm observing band.  Some of these are surprising; for instance, PSR~J1506$-$5158 was discovered in 1978 and modern-day instrumentation can easily obtain high S/N profiles for this pulsar with minimal integration time. We have therefore begun a project with the goal of completing the parameterisation of the pulsars in the catalogue and we have started with the missing flux density measurements. We note that, even though flux densities are fundamental properties of a pulsar, an in-depth study of pulsar emission or interstellar medium physics requires observations over wide-bandwidths and over different time scales (\citealt{ymc+18,slk+16}).    
%%% new paragraph for the referee report

An initial determination of a pulsar flux density can be obtained from the known survey sensitivity and the signal-to-noise of the detected profile.  Such flux density estimates have been published by e.g., \cite{mlc+01} and, more recently, \cite{slr+14}. After a pulsar has been discovered it is normally observed multiple times over at least 1\,yr.  During this time more accurate flux density determinations can be obtained (usually based on comparing the signal strength of the pulsar with that of a pulsed calibration source).  Pulsar data sets obtained with the Parkes radio telescope are usually placed on a flux density scale using observations of Hydra A (also known as 3C~218 and PKS~0915$-$11). This process assumes that Hydra A has a flux density of 43.1\,Jy at 1400\,MHz and a spectral index of $-$0.91 over the observation frequency range \citep{1977A&A....61...99B}. Observations are carried out by pointing directly at Hydra A and also recording data 1\,degree offset to the North and South.  During each observation the calibration signal is switched at 11.123\,Hz and these observations enable us to scale the measurement of the switched-calibration signal with a specific backend instrument to flux density units.  We then assume that the flux calibrator is stable between such measurements.   

Such calibration observations have been carried out over many years (with a cadence of approximately once per month) during Parkes Pulsar Timing Array (PPTA, \citealt{2013PASA...30...17M}) observations.  The data are processed and then made available for other observing projects.  In particular the numerous results from the PPTA have been combined together to give a small number of averaged flux calibration solutions and some of these are available online\footnote{{https://doi.org/10.4225/08/5ab2e8014512d}}.

In this paper, we:
\begin{itemize}
    \item Obtain new flux density measurements in the 20\,cm observing band for 32 pulsars.
    \item Compare flux density solutions obtained with Hydra A with those from another calibration source, PKS~B0407$-$658.
    \item Quantify the effects of radio-frequency interference (RFI), and its standard removal algorithms, on flux density determinations.
\end{itemize}

\section{Choosing the data sample and determining the flux densities}\label{sect:Obs}

The CSIRO data archive (see e.g., \citealt{2011PASA...28..202H}) contains the majority of pulsar observations carried out with the Parkes telescope (with the earliest observation in the archive from 1991). After an embargo period of 18 months the data become publically available.  We have therefore cross-matched the 20\,cm observations available in the archive with the pulsars that have no flux density measurements in the ATNF pulsar catalogue.  We have also managed to obtain a small number of extra observations with the Parkes telescope for a handful of pulsars.  In total we were able to obtain observations of 56 pulsars for which no flux densities were previously published in the 20\,cm band and have successfully obtained flux density values for 32 of these (the remainder were either undetectable or we only achieved a low S/N pulse profile). The pulsars in our sample are all young, solitary pulsars except for PSR~J1157$-$5112, which is in a binary system with a $\sim 3.5$\,day period and PSR~J1757$-$5322, which is a millisecond pulsar in a binary system with an orbital period of 0.5\,day.

All the observations that we have processed were obtained at a central frequency close to 1400\,MHz.  The majority were obtained with the central beam of the Parkes multibeam receiver (\citealt{1996PASA...13..243S}). A number of backend systems have been used for recording the data including the pulsar wide band correlator (WBC), the Parkes Digital Filterbanks (PDFB1, PDFB2, PDFB3 and PDFB4) and the CASPER Parkes Swinburne Recorder (CASPSR).  Details of receivers and backends can be found in  \citet{2013PASA...30...17M}.   All data were recorded using the PSRFITS data format (\citealt{2004PASA...21..302H}) with 30 or 60\,second sub-integrations.

The data were processed with the {\sc psrchive} pulsar signal processing system (\citealt{2004MNRAS.355..941H}). For our primary results, we excised data affected by narrow-band and impulsive RFI and removed 5\% of the band edges.  We used \textsc{pazi} to visually inspect the pulse profiles and to remove frequency channels or sub-integrations affected by RFI. We then used specific Hydra A flux calibration solutions corresponding to the receiver and backend instrument used \textcolor{cyan}. The majority of the calibration solutions were obtained from the data archive.  However, we reprocessed all the Parkes Pulsar Timing Array observations of Hydra A obtained with the multibeam receiver and PDFB4 backend system.  We produced both spot-measurements of the calibration solution and also produced a weighted-average of these calibration solutions providing an averaged solution for this receiver and backend system.  

The pulsar observations were calibrated using their associated calibration files using the {\sc psrchive} program {\sc pac} to flatten the bandpass, transform the polarization products to Stokes parameters and to calibrate the pulse profiles in flux density units. We formed analytic templates from our observations using \textsc{paas} and then obtained the flux density estimate using \textsc{psrflux}, which first matches the template with the observation and then determines the area under the template. This is carried out using a simple least-squares-fit in which the measured profile is assumed to be a scaled version of the template with an offset (Demorest, private communication). The uncertainty comes from the least squares uncertainty determination assuming that the reduced $\chi^2$-value of the fit is unity. Of course this procedure assumes that the template is a good representation of the pulse profile (we discuss below whether this is a reasonable assumption, or not). The observations for pulsars for which we have multiple observations were first processed independently, but then combined using \textsc{psradd} and the flux density procedure repeated to form a single, averaged, flux density value.

For comparison we also produced data files without any RFI-flagging and files with the automatic \textsc{paz} flagging, but without the manual RFI removal. We also obtained, for each of our data files, flux density estimates using non-averaged Hydra A flux calibration solutions and, in addition, flux calibration solutions that we have derived from observations of PKS\,B0407$-$658 (assuming a flux density of 16.02\,Jy at 1400\,MHz; \citealt{1981A&AS...45..367K}).  We have also trialled other procedures to measure the flux density of each pulsar including simply calculating the area under each profile without any assumption of the profile shape.

\section{Results and discussion}

\begin{table}
\caption{Pulsar name, pulse period, dispersion measure, observation length, signal-to-noise ratio for the folded profile and the flux densities for the pulsars in our sample. 1$\sigma$ uncertainties on the last quoted digit are given in parenthesis.}\label{flux}
\begin{tabular}{lllllll}
\hline
Psr \# & PSR J & Period & DM & Length & S/N & Flux density \\
      & & (s)    & (cm$^{-3}$pc) & (min) & & (mJy) \\
\hline
1 & J0133$-$6957 & 0.46 & 22.9 & 14.0 & 34.5 & 0.361(12) \\
2 & J0455$-$6951 & 0.32 & 94.9 & 77.0 & 11.1 & 0.083(4) \\
3 & J1006$-$6311 & 0.84 & 196.0 & 30.0 & 13.7 & 0.119(11) \\
4 & J1012$-$2337 & 2.52 & 22.5 & 15.0 & 8.2 & 0.135(13) \\
5 & J1057$-$4754 & 0.63 & 60.0 & 5.0 & 7.3 & 0.53(3) \\
\\
6 & J1157$-$5112 & 0.04 & 39.7 & 40.0 & 22.5 & 0.276(16) \\
7 & J1232$-$4742 & 1.87 & 26.0 & 5.0 & 43.6 & 2.38(6) \\
8 & J1312$-$5402 & 0.73 & 133.0 & 6.0 & 29.8 & 0.78(3) \\
9 & J1312$-$5516 & 0.85 & 134.1 & 3.5 & 119.4 & 3.04(3) \\
10 & J1328$-$4921 & 1.48 & 118.0 & 6.0 & 26.3 & 0.82(3) \\
\\
11 & J1335$-$3642 & 0.40 & 41.7 & 6.5 & 4.9 & 0.27(4) \\
12 & J1350$-$5115 & 0.30 & 90.4 & 2.5 & 51.0 & 1.27(3) \\
13 & J1355$-$5153 & 0.64 & 112.1 & 17.5 & 106.5 & 0.855(9) \\
14 & J1358$-$2533 & 0.91 & 31.3 & 3.0 & 4.9 & 0.15(5) \\
15 & J1414$-$6802 & 4.63 & 153.5 & 5.9 & 28.4 & 0.65(3) \\
\\
16 & J1420$-$5416 & 0.94 & 129.6 & 3.0 & 28.5 & 0.79(3) \\
17 & J1423$-$6953 & 0.33 & 124.0 & 15.0 & 24.2 & 0.362(9) \\
18 & J1457$-$5122 & 1.75 & 37.0 & 20.0 & 102.0 & 1.382(16) \\
19 & J1506$-$5158 & 0.84 & 61.0 & 2.0 & 60.4 & 3.75(8) \\
20 & J1603$-$2531 & 0.28 & 53.8 & 3.5 & 196.3 & 4.98(3) \\
\\
21 & J1610$-$1322 & 1.02 & 49.1 & 30.0 & 56.6 & 1.11(3) \\
22 & J1659$-$1305 & 0.64 & 60.4 & 30.0 & 47.2 & 0.803(17) \\
23 & J1708$-$7539 & 1.19 & 37.0 & 30.0 & 52.6 & 0.747(13) \\
24 & J1711$-$5350 & 0.90 & 106.1 & 6.0 & 39.9 & 0.836(19) \\
25 & J1728$-$0007 & 0.39 & 41.1 & 30.0 & 46.0 & 0.655(15) \\
\\
26 & J1734$-$0212 & 0.84 & 65.0 & 40.0 & 26.6 & 0.279(12) \\
27 & J1749$-$5605 & 1.33 & 58.0 & 5.5 & 22.7 & 0.635(15) \\
28 & J1757$-$5322 & 0.01 & 30.8 & 20.0 & 37.3 & 1.17(4) \\
29 & J1833$-$6023 & 1.89 & 35.0 & 29.9 & 123.1 & 1.459(12) \\
30 & J1857$-$1027 & 3.69 & 108.9 & 28.3 & 164.1 & 2.032(17) \\
\\
31 & J1900$-$7951 & 1.28 & 39.0 & 27.9 & 77.4 & 0.946(12) \\
32 & J2155$-$5641 & 1.37 & 14.0 & 39.9 & 31.4 & 0.361(12) \\
\hline
\end{tabular}
\end{table}

Our primary results are listed in Table~1.  In column order we give each pulsar an identification number and then list the pulsar name, pulse period, dispersion measure, observation length, signal-to-noise of the pulse profile and the measured flux density obtained using the most up-to-date Hydra A calibration solution. We recommend that these flux density values be included in the next version of the pulsar catalogue.   In general, our results are not surprising. 
\cite{jk+18} published flux densities for about 600 pulsars at 1.4 GHz with Parkes archive data. Most pulsars in their paper have multiple observations and they add all the data together and present a single flux density for a given pulsar. In order to compare the results from our pipeline with theirs for a given observation, we choose two pulsars (PSRs~J0540$-$7125 and J0758$-$1528; a weak and a moderately-weak pulsar) with only one observation. We find the flux densities obtained with our pipeline is consistent with theirs at a level of 0.1\,mJy. The differences could be caused by the various factors we discuss later in this section. We note a strong correlation between the profile S/N and the flux density and a strong inverse correlation between the profile S/N and the fractional size of the uncertainty.  For the remainder of this paper we wish to determine how confident we can be in these flux density values and their uncertainties.

\subsection{The Hydra A calibration solutions}

Our flux density values rely on the assumed flux density (and its spectral index) for the radio galaxy Hydra A. The flux density and spectral index at 1400\,MHz of Hydra A have been published by different authors (\citealt{1977A&A....61...99B,1981A&AS...45..367K,2004AJ....127...48L}) and range from $\sim$43 to 45\,Jy. These discrepancies in these published measurements, could, as noted by \citet{2018MNRAS.473.4436J}, be caused by different pointing positions along the two radio lobes and are much larger than the formal uncertainties on most of our flux density values.  The ratio of these values is 0.96 implying an uncertain scaling factor in any measured flux density value of this factor.

We can also compare our results to those obtained using PKS~B0407$-$658 as the primary calibrator.  Over a narrow bandwidth, changing the primary calibrator should simply scale the resulting flux density values by a small factor and we find that our nominal results listed in Table~1 scale by a mean factor of 0.914 (with a standard deviation of 0.001) when we make use of PKS~B0407$-$658 (i.e., the flux density values are slightly smaller when we use PKS~B0407$-$658 than when we use Hydra A).  As we have no prior reason to believe one flux calibrator over the other, we must assume that there is an uncertainty in our measurements of around this factor caused by the choice of flux calibration solution. 

\begin{figure} 
\centering
 \includegraphics[width=14cm, angle=0]{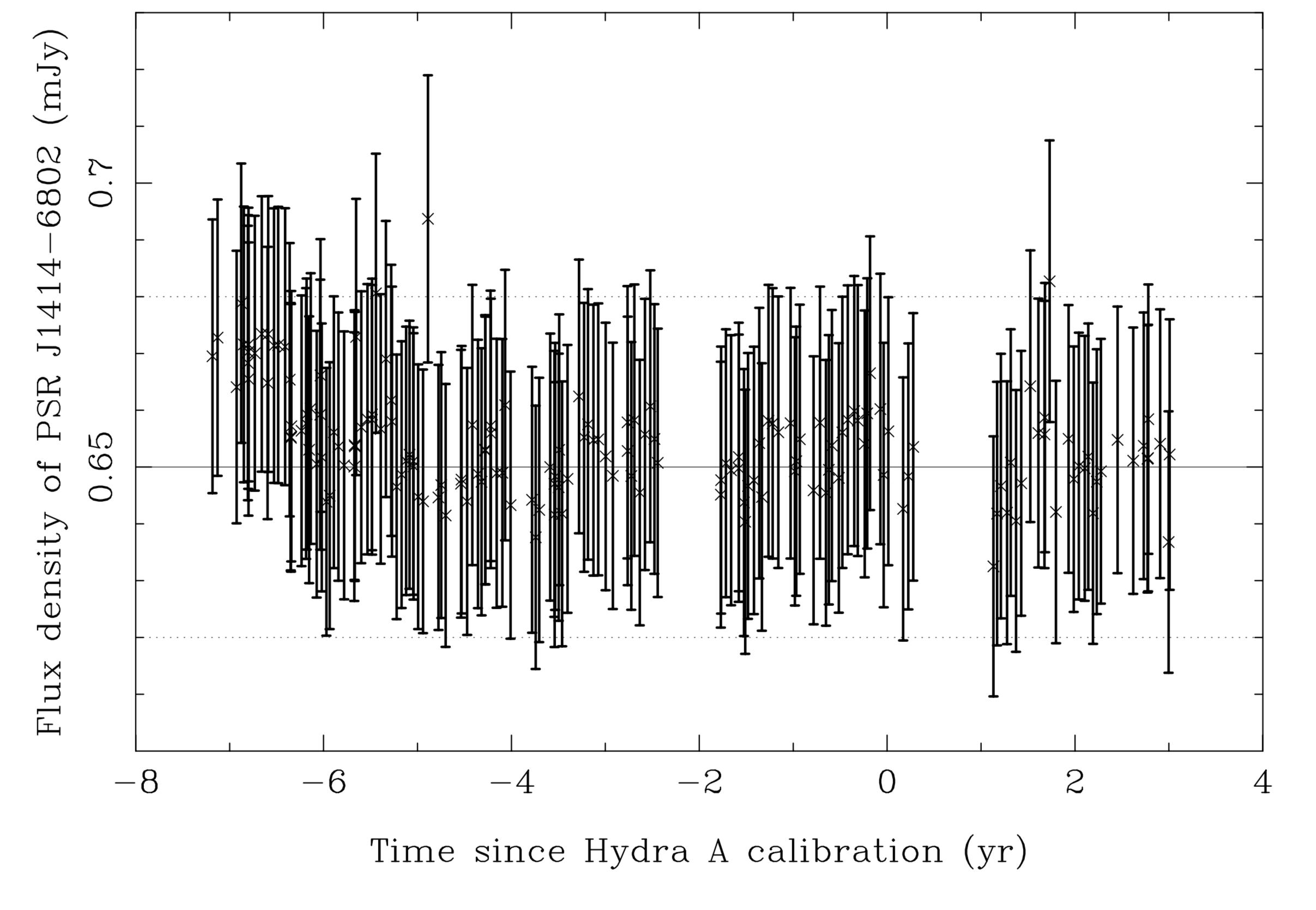}
 \caption{The flux density of PSR~J1414$-$6802 when determined using different Hydra A calibration solutions. The solid horizontal line and the dotted lines indicate the flux density of this pulsar as reported in Table~\ref{flux}.}
 \label{testFluxCal}
\end{figure}

We can also trial different flux calibrator solutions derived from different observations of Hydra A.  We have chosen to study PSR~J1414$-$6802 for which we have a recent observation and a measured flux density in Table~\ref{flux} of 0.66(3)\,mJy.  We have trialled 168 different Hydra A calibration solutions measured over the last decade. All of these Hydra A calibration observations were carried out using the central beam of multibeam receiver and PDFB4 backend at central frequecy of 1369~MHz with 256~MHz bandwidth.
%% the first two points in Figure~1 were observed with PDFB3, should we remove them or use different color?
%%%
Our results are shown in Figure~\ref{testFluxCal} as a function of the time since the Hydra A calibration observation was made.  There is a variation in the measured flux density between different calibration solutions.  The standard deviation of the flux density values is 0.0094\,mJy suggesting that a single flux density solution will only provide a flux density with a precision of $\sim$0.01\,mJy\footnote{Note that there is no evidence for any correlation between the measured flux density and the pointing position of the telescope during the flux calibration method.} (a detailed description of the flux density measurements and correlations with changes in the observing system will be described in the forthcoming PPTA second data release publication).

\subsection{Measuring the flux densities}

\begin{figure} 
\centering
 \includegraphics[width=14cm, angle=0]{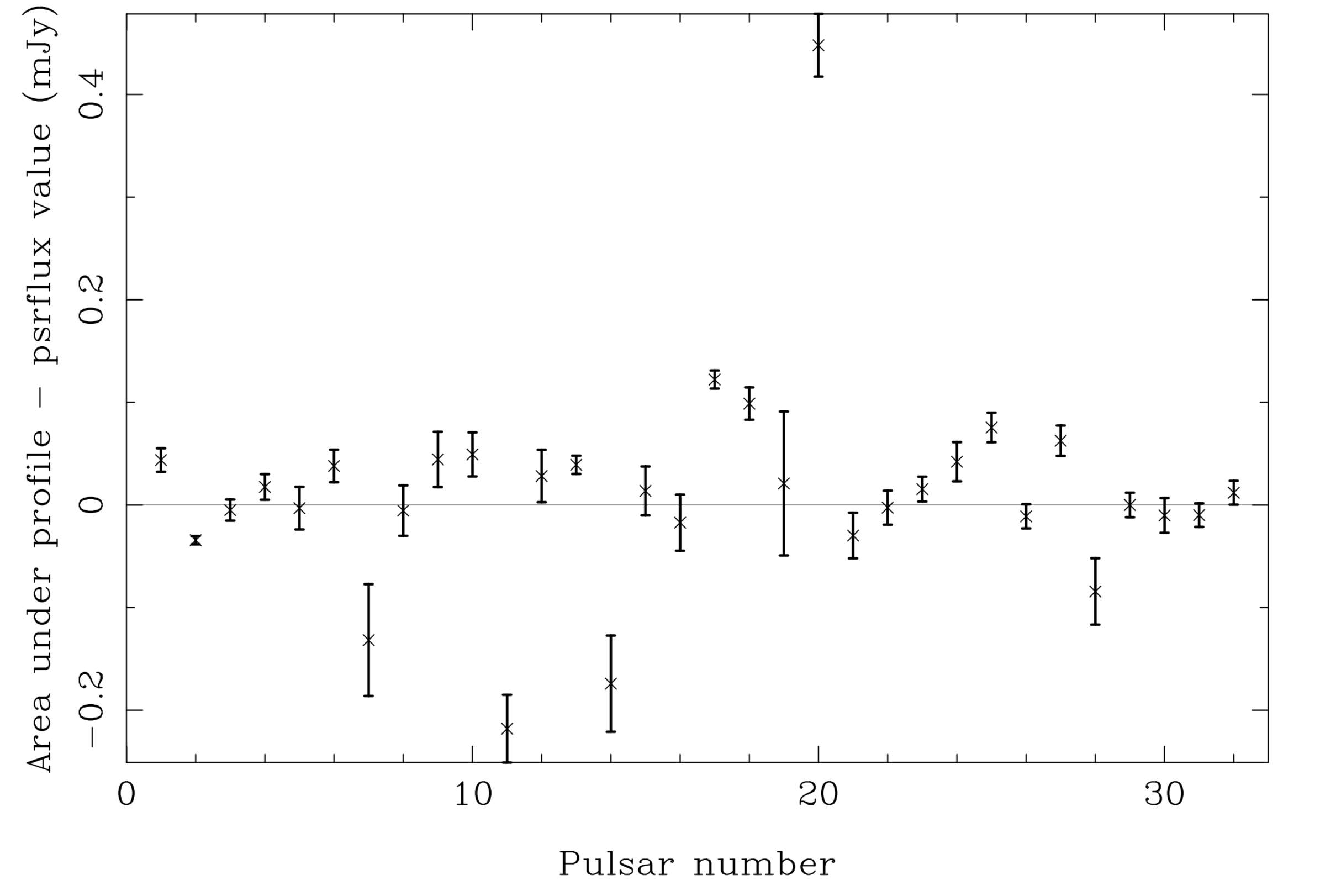}
 \caption{The flux density measured for each pulsar by simply summing under the profile (with a baseline removed) versus the determination using an analytic pulse template.}
 \label{fg:plotArea}
\end{figure}

The flux density values are, of course, dependent upon the method used to make the measurement. Published flux density values represent the signal strength that would be recorded for the point continuum source at the position of the pulsar (i.e., it is determined across both the on- and off-pulse emission).  This implies that an estimate of the flux density can simply be made by summing the calibrated pulse profile across all the pulse phase bins (assuming that a baseline has been removed from the off-pulse emission).  In Figure~\ref{fg:plotArea} we plot the difference between the flux density obtained by summing the profile bins minus our nominal flux densities (the error bars are taken as uncertainties on the values in Table~\ref{flux}). Simply summing under the profile will be biased high because of the presence of RFI and, for most of the pulsars, we note higher values of $\sim 0.05$\,mJy on average. However, the procedure that makes use of a pulse profile will be affected by any non-perfect match between the analytic template and the actual profile. As small-scale, narrow features are hard to model using existing software then the template procedure is liable to be biased low.  For our sample of pulsars PSRs~J1457-5122, J1749$-$5605 and J1857$-$1027 have two clear components in the pulse profile.  For these pulsars we have created two standard templates. One has only a single component, whereas the other has a two or more components. The measured flux densities have, as expected, significantly smaller uncertainties when the template matches the profile and the measured values changes by 0.1, $-$0.25 and 0.1\,mJy respectively for the three pulsars.

\subsection{Removing RFI}

\begin{figure} 
\centering
 \includegraphics[width=14cm, angle=0]{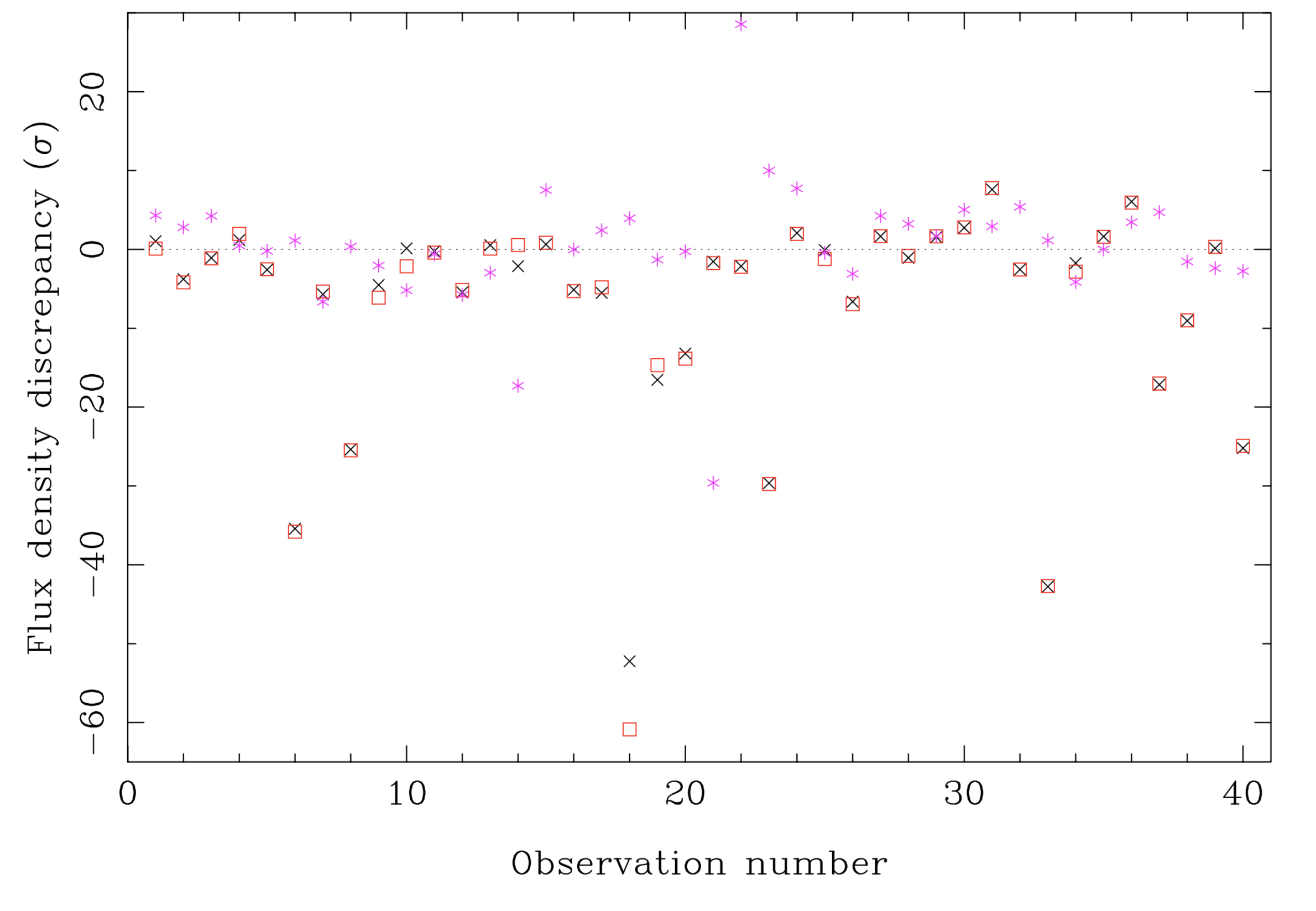}
 \caption{A comparison between the flux density measurements without any RFI removal (red squares) and with a basic, automated scheme (black crosses) compared with those from manual RFI zapping. The purple points indicate the difference in the measured flux densities when two different people carry out their own manual RFI removal. }
 \label{fg:compareRFI}
\end{figure}

All the methods described above will be affected by RFI present in the profile.  However, manual RFI removal is time intensive, difficult to reproduce at a later date, and is challenging as we move into the era of very large data sets. We have therefore compared how our flux density measurements vary between the nominal results (listed in Table~\ref{flux}), which had both automatic and manual RFI zapping, with profiles obtained with (1) no RFI removal at all and (2) only the automated RFI removal (in which the PSRCHIVE package \textsc{paz} is used, with default parameters, to remove frequency channels containing significant RFI using a median-zapping routine).  Our results are provided in Figure~\ref{fg:compareRFI}.  We note that the flux densities measured without manual RFI flagging are usually larger (because the presence of RFI will positively bias the results)\footnote{In a few cases the flux densities are smaller without manual RFI flagging. This is because some of the actual pulsed signal was removed during the flagging process.} and that the flux densities can vary by much more than the uncertainties on the flux density measurements.  The differences between no RFI flagging at all, and a simple automated method are slight. 

Any manual RFI removal will depend on the person carrying out the zapping. We therefore also compare the results provided in Table~\ref{flux} with the same measurements made by a different co-author on our paper (note that this involves variations in the manual RFI flagging and in the manual determination of the pulse template). In Figure~\ref{fg:compareRFI} we compare the results given in Table~\ref{flux} with those determined independently (blue point).  The root-mean-square deviation is 0.11\,mJy. We conclude that current procedures that require no human interaction and are therefore reproducible and quick, lead to significant variations in the measured flux density values. However, we also note that different individuals carrying out manual RFI flagging procedures will also lead to variations in the measured flux densities.

\subsection{Spectral indices}

We combine our flux density measurements with published values obtained in different observing bands (close to 300, 600, 700 and 3000\,MHz). These are listed in Table~\ref{index} as S300, S600, S700 and S3000, respectively.  The S1400 column represents our measurement at 1400\,MHz. We can use these multi-frequency observations to estimate each pulsar's spectral index, $\alpha$, assuming a simple scaling law $S_\nu = bx^\alpha$, where , $S_\nu$ is estimated total flux for a given frequency, $\nu$. Values of $b$ and $\alpha$ are given in the Table. Most of these spectral indices are steeper than the mean values from \citet{2018MNRAS.473.4436J}, but our measurements are not particularly unusual as many other pulsars do have similar spectral indices.  We do not believe that our pulsar sample is particularly biased towards steep spectrum pulsars and therefore  this result is simply because of our relatively small sample.

\begin{table}
\caption[]{Flux densities and spectral indices for nine pulsars where multi-frequency flux densities are available \label{index}}
\setlength{\tabcolsep}{1pt}
\small
\begin{tabular}[H]{p{1.8cm}p{1.2cm}p{1.3cm}p{1.3cm}p{1.6cm}p{1.4cm}p{1.5cm}p{1.5cm}}
\hline\noalign{\smallskip}
PSR J& S300& S600& S700& S1400& S3000& b& $\alpha$\\
&(mJy)& (mJy)& (mJy)& (mJy)&(mJy)&&\\
\hline\noalign{\smallskip}
J1012$-$2337&       &1.3&     &0.135(13)&       &    0.17& $-$2.66\\
J1312$-$5402&  41(1)&	     & 5(1)&  0.78(3)&0.13(4)& 1.06(2)& $-$2.49(1)\\
J1312$-$5516&       & 17&12(3)&  3.04(3)& 0.6(2)&3.39(20)& $-$2.07(82)\\
J1355$-$5153&  50(5)&	     &	   & 0.855(9)&		 &    1.05& $-$2.63\\
J1610$-$1322&	      &  7(1)&	   &  1.11(3)&		 &    1.30& $-$2.17	\\
J1659$-$1305&	      &3.2(3)&	   &0.803(17)&		 &    0.91& $-$1.63	\\
J1711$-$5350&       &	     & 4(1)&0.836(20)&0.19(6)&1.01(33)& $-$2.22(54)\\
J1728$-$0007&	      &4.1(6)&	   &0.655(15)&		 &    0.77& $-$2.16\\
J1900$-$7951&	      &  5& 4(2)&0.946(12)&       &1.21(21)& $-$1.86(24)\\
\noalign{\smallskip}\hline
\end{tabular}
\label{tab:pulsars}
\end{table}

\section{Conclusion}

In this paper, we have presented the flux densities at 1.4\,GHz for 32 pulsars using the Parkes radio telescope.   We have shown that the procedure used in processing the data and calculating the flux densities will affect the final result usually at a level greater than the formal uncertainties published. 

In summary:
\begin{itemize}
    \item Hydra A, which is commonly used as the primary flux calibration, has published flux densities that vary by a factor of 0.96 at 1.4\,GHz.
    \item Different measurements of Hydra A lead to random variations in the measured flux density values of $\sim 0.01$\,mJy.
    \item Using PKS B0407$-$658 as the primary flux calibrator instead of Hydra A scales the flux density values by a factor of 0.91.
    \item Various methods to measure a pulsar's flux density from its profile have been proposed. The differing methods lead to changes in the measured flux densities of $\sim 0.05$\,mJy, which, in some cases, is larger than the formal measurement uncertainty.
    \item RFI removal schemes can have a large effect on the resulting flux density values (leading to variations of many mJy).  
\end{itemize}

 The largest factor in flux density estimation relates to RFI removal. With the existing systems at Parkes, and with the use of Hydra A as the primary flux density calibrator, it is unlikely that a given flux density measurement will be more precise than $\sim$0.01\,mJy.

A spot measurement of a pulsar spectral index has limited use as pulsar flux densities vary because of diffractive and refractive scintillation (as well as intrinsic flux variations), but with straight-forward calibration schemes it is possible to obtain flux calibrated data sets over many epochs.   We will continue this work in the future using both the Parkes and FAST telescopes and we note that there are a large number of pulsars in the catalogue for which we still do not have such basic information as the flux density in the 20\,cm band.  For instance, there are 49 pulsars detected in early Parkes-telescope surveys with no measured flux density measurements and a further four discovered by the Molonglo telescope.  The new Ultra Wide-bandwidth Low receiver on Parkes will be ideal for this type of analysis as it is highly sensitive and covers a wide observing band. However, the pulsar catalogue will need to be updated as, currently for each pulsar, a single flux density measurement is recorded in well-defined observing bands.  As pulsar flux densities vary in time and frequency, we recommend that the catalogue be updated to store both time and frequency information for each flux density measurement for each pulsar. 

The Parkes Pulsar Timing Array project will continue to obtain flux density calibration solutions for the general user community.  The choice of Hydra A is not ideal, in part due to the source having extended structure and is not carefully monitored by any observatory. An ideal calibrator would be a bright point source, which has constant flux, and is located in an empty sky field; however, such a target is unrealistic. Unresolved, extragalactic sources typically have stable fluxes (with the noted exception of quasars) and variations in flux densities between observatories can often be attributed to the varying number of external sources in the field. Using an interferometer in conjunction with single-dish observations of a flux calibrator can help to alleviate any discrepancies by sampling with a smaller synthesis beam and imaging potentially complex fields. One such potential flux calibrator candidate is PKS~2251+158 (3C454.3), which is located at a declination accessible to almost all radio observatories (including both FAST and Parkes).

Calibrating radio astronomy data sets is both essential and non-trivial.  With complex telescope systems (such as FAST, in which the surface of the dish and the focus cabin position continuously moves during a tracking observation) and new ultra-wide-band, or wide field-of-view, receivers new calibration procedures are becoming necessary. Novel calibration methods such as switching calibration signals at extremely high rates (Li, private communication), injecting pseudo-random-noise sequences, or short duration pulses (\citealt{2017PNAS..114E9188P} are all currently being explored and we look forward to comparing these methods with the traditional calibration procedures in the near future.  

\normalem

\begin{acknowledgements}
The Parkes radio telescope is part of the Australia Telescope National Facility which is funded by the Commonwealth of Australia for operation as a National Facility managed by CSIRO. This paper includes archived data obtained through the CSIRO Data Access Portal (http://data.csiro.au). This work is supported by the Youth Innovation Promotion Association of Chinese Academy of Sciences, 201$^*$ Project of Xinjiang Uygur Autonomous Region of China for Flexibly Fetching in Upscale Talents, the National Key R\&D Program of China (No. 2017YFA0402602) and the Strategic Priority Research Programme (B) of the Chinese Academy of Sciences (No. XDB23010200).The FAST FELLOWSHIP is supported by Special Funding for Advanced Users, budgeted and administrated by the Center for Astronomical Mega-Science, Chinese Academy of Sciences. We thank the Parkes Pulsar Timing Array project team for allowing us to use their unpublished Hydra A calibration solutions.

\end{acknowledgements}
\bibliographystyle{raa}
\bibliography{bibtex}
\end{document}